\documentclass[aps,pra,reprint,superscriptaddress]{revtex4-1}

\usepackage{SIunits}
\usepackage{graphicx}
\usepackage{xcolor}
\usepackage{appendix}
\usepackage{chemformula}

\begin{document}

\title{Cavity Optomechanics with Photonic Bound States in the Continuum}

\author{Jamie M. Fitzgerald}
 \email{jamief@chalmers.se}
 \affiliation{Department of Physics, Chalmers University of Technology, G\"oteborg, Sweden}
\author{Sushanth Kini Manjeshwar}
\affiliation{Department of Microtechnology and Nanoscience, Chalmers University of Technology, G\"oteborg, Sweden}
\author{Witlef Wieczorek}
\affiliation{Department of Microtechnology and Nanoscience, Chalmers University of Technology, G\"oteborg, Sweden}
\author{Philippe Tassin}
\affiliation{Department of Physics, Chalmers University of Technology, G\"oteborg, Sweden}

\date{\today}

\begin{abstract}
We propose a versatile, free-space cavity optomechanics platform built from two photonic crystal membranes, one of which is freely suspended, and designed to form a microcavity less than one wavelength long. This cavity features a series of photonic bound states in the continuum that, in principle, trap light forever and can be favorably used together with evanescent coupling for realizing various types of optomechanical couplings, such as linear or quadratic coupling of either dispersive or dissipative type, by tuning the photonic crystal patterning and cavity length. Crucially, this platform allows for a quantum cooperativity exceeding unity in the ultrastrong single-photon coupling regime, surpassing the performance of conventional Fabry-P\'{e}rot-based cavity optomechanical devices in the non-resolved sideband regime. This conceptually novel platform allows for exploring new regimes of the optomechanical interaction, in particular in the framework of pulsed and single-photon optomechanics.
\end{abstract}

\maketitle

Cavity optomechanical devices~\cite{Aspelmeyer2014} provide quantum control over their constituent mechanical and optical degrees of freedom for use in precision measurements, quantum networks, and fundamental tests. To this end, optomechanical devices require sufficiently strongly coupled optical and mechanical resonators, along with the minimization of unavoidable decoherence, so they can access the strong-cooperativity regime. Experiments have accessed this regime by boosting the optomechanical interaction with a laser drive, resulting in the demonstration of ground-state cooling~\cite{Chan2011,Teufel2011,Rossi2018,Delic2020}, optical~\cite{Safavi2013,Purdy2013} or mechanical squeezing~\cite{Wollman2015,Pirkkalainen2015,Lecocq2015}, or (opto)mechanical entanglement~\cite{Palomaki2013,Ockeloen2018,Riedinger2018}. These experiments have exploited a linear coupling to the mechanical resonator, while nonlinear coupling enables complementary ways to measure and manipulate mechanical motion in the quantum regime~\cite{Thompson2008,Sankey2010,clerk2010quantum,purdy2010tunable,vanner2011selective,Paraiso2015}. 
 
 Cavity optomechanical platforms can be classified based on whether in-plane or out-of-plane light propagation is used. While in-plane geometries boast the largest coupling rates due to co-localization of photonic and phononic modes~\cite{Chan2011,Gavartin2011,Leijssen2017,guo2019feedback}, they are inherently limited by material loss and structural disorder. The advantage of out-of-plane geometries, such as Fabry-P\'erot (FP) cavities in end-mirror~\cite{Gigan2006,Arcizet2006,Kleckner2006}, membrane in the middle (MiM) ~\cite{Thompson2008}, or levitated~\cite{Delic2020} configurations, is that a substantial proportion of light propagation is in vacuum. This leads to a lower optical decay rate, but comes at the price of smaller single-photon coupling rates. In part, the original motivation for the MiM setup was to spatially separate the mechanical and optical functionality, but the resulting weak coupling has naturally led to attempts to increase it. In particular, the concept of multi-element optomechanics \cite{Xuereb2012} has been proposed, but its realization is involved~\cite{Nair2017,Piergentili2018,Gartner2018,Wei2019}. Furthermore, as light is trapped in ever smaller volumes, the necessity of the outer cavity becomes questionable~\cite{Xuereb2013}. 

\begin{figure}
\includegraphics[width=8.5cm]{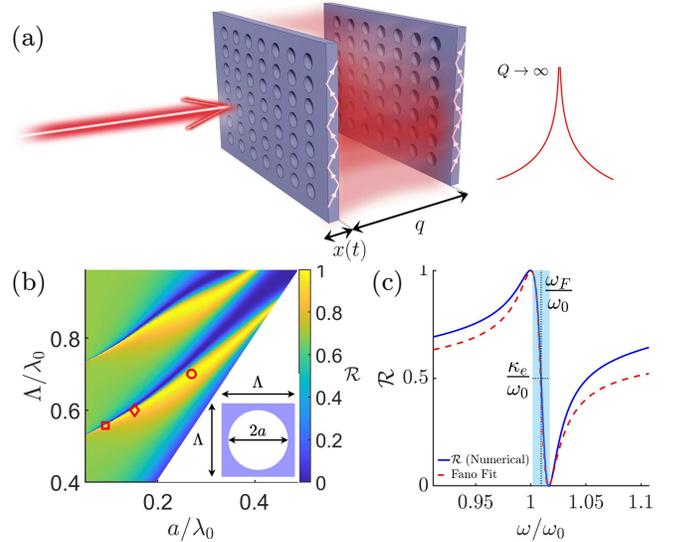}
\caption{\label{fig:fig1} (a)~Illustration of the double photonic crystal slab cavity (DPhoC) and the bound state in the continuum (BIC) mechanism. (b)~Reflectance map of lattice period and air-hole radius for a \unit{100}{\nano\meter}-thick GaAs slab designed to operate at a wavelength $\lambda_0 = \unit{1550}{\nano\meter}$. The red markers indicate the three parameter sets used in this Letter. (c)~Reflectance spectrum of a PhC slab (blue line) with $\Lambda=0.6 \lambda_0$ and $a=0.1525\lambda_0$, and an exemplary Fano fit (red-dashed line).}
\end{figure}

The need in optomechanics for high reflectivity, high mechanical quality factor, and low mass mechanical resonators necessitates a move away from bulky components such as Bragg mirrors, towards ultra-thin mirrors. Suspended photonic crystal (PhC) slabs that support guided-mode resonances~\cite{Fan2003} have been demonstrated to posses over 99.9\% reflectance~\cite{Chen2017} without compromising on the mechanical properties~\cite{Norte2016,Gartner2018,Manjeshwar2020}. In reflection and transmission spectra, the guided mode manifests as an asymmetrical Fano lineshape~\cite{Fan2003}. Placing two PhC slabs close together has long been considered for sensing applications~\cite{Suh2005,Shuai2013,Liu2017}, and experimental studies have explored placing a single PhC slab in a cavity~\cite{Kemiktarak2012,Woolf2013,Makles20152,Chen2017,Stambaugh2015}, as well as two PhC slabs in a MiM configuration~\cite{Gartner2018} and as a cavity in their own right~\cite{Roh2010}. It has only recently become apparent that the internal dynamics of the guided-mode resonance can lead to new optomechanical effects~\cite{Naesby2018,Vcernotik2019}.

In this Letter, we propose a novel platform for cavity optomechanics, constructed from two suspended PhC slabs in an end-mirror configuration, that relies on photonic bound states in the continuum (BICs). BICs are a general wave phenomenon, where a completely spatially localized mode can exist above the light line~\cite{Hsu2016}. The double PhC slab cavity (DPhoC), depicted in Fig.~\ref{fig:fig1}(a), possesses a large optomechanical coupling due to its near-wavelength length and a moderately low decay rate thanks to the near-perfect trapping of light via the BIC. In contrast to conventional end-mirror systems, this simple system can access purely linear dispersive optomechanical coupling via the BIC mechanism, or purely quadratic coupling via evanescent coupling between the slabs. We argue that near-/sub-wavelength localization of optical modes is a promising strategy for out-of-plane systems and that the DPhoC, with experimentally realistic parameters, can simultaneously possess the required optical and mechanical properties to access the strong quantum cooperativity regime on the single-photon level, without the encumbrance of an outer FP cavity. 

\begin{figure}
\includegraphics[width=8.5cm]{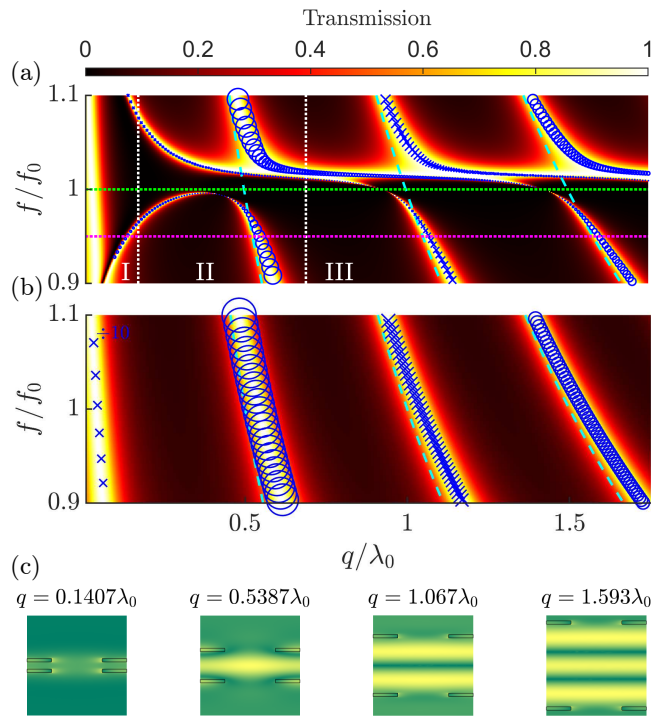}
\caption{\label{fig:fig2} (a)~Transmittance map against frequency and separation for two PhC slabs with a period of $0.6\lambda_0$ and radius of $0.1525\lambda_0$, with the cavity eigenmodes overlaid in blue. (b)~Transmittance map for two homogeneous slabs with an effective refractive index given by the lattice parameters of (a). (c)~Electric field plots for separations corresponding to peak transmittance taken along the slice $f=0.95 f_0$ for the DPhoC, as indicated by the magenta dotted line in (a).} 
\end{figure}

\emph{Optomechanical Couplings}.--- For a FP cavity with a movable end mirror, the cavity mode energy, $\hbar \omega_c$, depends parametrically on the resonator's out-of-plane displacement $x$ and can be expanded around the equilibrium point, $q$, leading to the \emph{linear}, $g_0= - {\partial \omega_c}/{\partial x}|_{x=q}\,x_{0} = G x_0$, and \emph{quadratic}, $g_2= - \frac{1}{2} {\partial^2 \omega_c}/{\partial x^2}|_{x=q}\,x_{0}^2 = G_2 x_0^2$, single-photon coupling rates, where $x_{0}$ is the zero-point motion. We have introduced for convenience the optical frequency shift per displacement, $G$, and its counterpart for the second derivative, $G_2$. Both $g_0$ and $g_2$ are complex numbers as $\omega_c$ is the eigenvalue of an open-cavity problem; the imaginary part gives the decay of the cavity mode, $\kappa$. Thus, the real part of the coupling describes dispersive coupling and the imaginary part describes dissipative coupling.

\emph{The Double Photonic Crystal Slab Cavity}.--- Inspired by our recent experimental work~\cite{Manjeshwar2020}, the model system is built from \unit{100}{\nano\meter}-thick, GaAs PhC slabs patterned with a square lattice of circular holes, and designed to operate at a wavelength around $\lambda_0 = \unit{1.55}{\micro\meter}$, where GaAs has a high refractive index of $3.374$~\cite{Guha2017}. We stress at this point that the physics discussed in this work is not material dependent (see also appendix \ref{appendix_6}) and we expect the same phenomena in, e.g., SiN-based systems~\cite{Norte2016,Bernard2016,Chen2017,Nair2017,Piergentili2018,Gartner2018}. To find suitable lattice parameters to achieve a Fano resonance for a single slab, a reflectance map over the air-hole radius and lattice period is calculated. See appendix \ref{appendix_1} for details on the numerical calculations. The lattice parameters used in this section are indicated by the red diamond marker on Fig.~\ref{fig:fig1}(b): a period of $0.6\lambda_0$ and radius of $0.1525 \lambda_0$. The reflectance spectrum, depicted in Fig.~\ref{fig:fig1}(c), shows a pronounced peak near $\lambda_0$, which corresponds to the Fano resonance. The physics behind this is well captured by coupled-mode theory~\cite{Haus1984,Fan2003} and an exemplary Fano fit is plotted in Fig.~\ref{fig:fig1}(c), which allows for the extraction of the mode frequency $\omega_F$, external decay rate (radiative loss) $\kappa_e$, and internal decay rate $\kappa_i$ (e.g., due to materials loss). More details are given in appendix \ref{appendix_2}.

We now consider two such PhC slabs, separated by a distance $q$. The transmittance spectrum is mapped for a range of separations, which is shown in Fig.~\ref{fig:fig2}(a). To emphasize that the coupled PhC slabs do not simply lead to a higher reflectivity, but rather new phenomena, we show in Fig.~\ref{fig:fig2}(b) the transmittance map of the corresponding double homogeneous-slab system with an effective refractive index (see appendix \ref{appendix_2})~\cite{Fan2003}. On top of both transmittance maps, the electromagnetic eigenfrequencies are shown in blue, which are sorted by their even (crosses) or odd (circles) symmetry. The imaginary part of the eigenfrequency measures the radiative loss of the mode and its value is indicated by the marker size. At this stage, no material loss is included.

To aid our discussion, we identify three regions based on the slab separation and the consequent dominant form of interaction between the slabs: near field ($q\lesssim  \lambda_0/(2 n_{\mathrm{eff}}) \sim \unit{270}{\nano\meter}$), intermediate field ($ \lambda_0/(2 n_{\mathrm{eff}}) \lesssim  q \lesssim  2 \lambda_0/n_{\mathrm{eff}}$), and far field ($q \gtrsim 2 \lambda_0/n_{\mathrm{eff}} \sim \unit{1000}{\nano\meter}$), separated by white lines in Fig.~\ref{fig:fig2}(a). In the far-field, the transmittance maps shown in Figs.~\ref{fig:fig2}(a) and (b) exhibit diagonal bands of high transmittance, which is typical FP behaviour. The cavity mode energies of a perfect FP cavity are indicated by the dashed cyan lines. For a given frequency, the transmittance reaches unity for certain separations where a half-integer number of wavelengths can fit into the cavity [see Fig.~\ref{fig:fig2}(c)]. In contrast to the homogeneous slabs, which closely follow the cyan lines, the structured slabs show a much more intricate structure on top of this background. Most pertinent to our discussion is the narrowing of the transmittance bands close to the Fano resonance (indicated by the green-dotted line). Further inspection shows that the linewidth of the transmittance spectra approaches zero close to separations corresponding to FP resonances. This behavior is captured by the marker size of the eigenmodes becoming vanishingly small. This is fundamentally different behaviour from a conventional FP cavity where the decay rate is inversely proportional to the cavity length; instead, it is indicative of the evolution of the cavity eigenmode into a BIC.\\
\indent BICs are peculiar resonances that do not decay over time as there are no available radiation channels due to destructive interference; in principle, they have an infinite quality (Q) factor~\cite{Hsu2016}. They have been explored in PhC slabs~\cite{Hsu2013,Gansch2016,Kodigala2017} and double PhC slab structures~\cite{Li2016}. In practice, the optical Q-factor is limited by structural disorder and material loss; nevertheless, Q-factors up to $4.9\times 10^5$ have recently been demonstrated~\cite{Jin2019}. The BICs we observe here are examples of ``resonance-trapped'' BICs, where the gap acts as the tunable parameter~\cite{Hsu2016}, which are attractive as they are quite robust to imperfections---one need only change the tuning parameter to compensate for geometrical perturbations. An infinite number of BICs exist for the DPhoC for increasing $q$, but occur at ever smaller gradients, indicating a weaker optomechanical coupling. The long lifetime of the photon in the guided-mode resonance allows a moderately low decay rate even for wavelength-sized cavities. This observation is extremely relevant for microcavities for optomechanics; we can boost $g_0$ by reducing the cavity length down to $\sim \lambda_0/2$, but with $\kappa$ not limited by the cavity length but rather by the internal loss of the individual slab resonances. A comparison between conventional FP-type optomechanical microcavities and the DPhoC is presented in appendix \ref{appendix_5}. Despite the huge amount of current interest in BICs~\cite{Hsu2016}, there has been limited study of their utility for optomechanics~\cite{zhao2019mechanical,hurtado2020bound}. 

\begin{figure*}
\includegraphics[width=16cm]{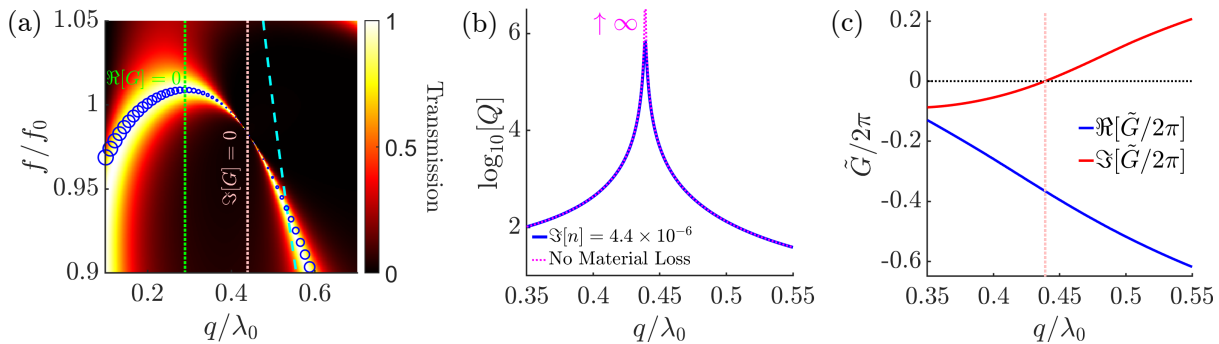}
\caption{\label{fig:fig3} (a)~Transmittance map against frequency and slab separation for a DPhoC with a lattice period of $0.7\lambda_0$ and air-hole radius of $0.27\lambda_0$. (b)~Quality factor of the cavity eigenmodes close to the BIC. (c)~Dispersive and dissipative parts of the normalized optical frequency shift per displacement: $\tilde{G}/2\pi=\frac{d f_c/f_0}{dq/\lambda_0}$. The BIC is indicated by the dotted pink line.} 
\end{figure*}
   
The lowest-order BIC is located in the intermediate field, where both coupling via photon tunneling, associated with gradient forces, and propagation, associated with radiation pressure, are relevant. This is illustrated in the electric field plot for $q=0.5387\lambda_0$ in Fig.~\ref{fig:fig2}(c) by the deviation from the standard standing-wave mode profile. In this region, we see the same linewidth narrowing, but now the high-transmittance band is highly warped and bends away from the FP line. We also observe the very typical mode splitting of an odd and even mode around the Fano energy (see appendix \ref{appendix_3}). Furthermore, for the lower-energy even mode, there is a crossover from a repulsive to attractive force, i.e., at a certain separation the derivative with respect to displacement vanishes, allowing the DPhoC to access purely quadratic optomechanical coupling: $g_0 = 0, g_2 \neq 0$. This is in stark contrast to the regular end-mirror configuration, which can only support repulsive forces. The quadratic coupling relies on gradient forces, which depend on the overlap between the near fields of both slabs, and so exhibits an exponential dependence on separation~\cite{Suh2005,Van2010}: $\zeta \propto \exp{(-q/\delta)}$, where $\delta$ quantifies the out-of-the-plane decay length of the guided mode. The use of evanescent coupling in optomechanics is nothing new; it has been commonly used to couple light in waveguides to optical microresonators~\cite{Eichenfield2007,Anetsberger2009}, as well as microresonators to one another~\cite{Wiederhecker2009}, but has rarely been utilized for out-of-plane optomechanics~\cite{Roh2010,Woolf2013}.

All of the physics displayed in the transmittance map in Fig.~\ref{fig:fig2}(a) can be captured extremely well by coupled-mode theory~\cite{Haus1984,Fan2003,Suh2005}. We  fit the expressions obtained from coupled-mode theory to the results of numerical simulations to find the value of $\zeta$ and find excellent agreement. More importantly, the theory provides an explanation for the family of BICs we observe. By ignoring the direct reflection and transmission of light through the slab, and considering only the interaction via the excited Fano resonances, we find that the BICs are a predominately far-field phenomena found close to the FP resonances where the cavity decay rate completely vanishes in the absence of internal loss. The details are given in appendix \ref{appendix_3}.

\begin{figure}
\includegraphics[width=8cm]{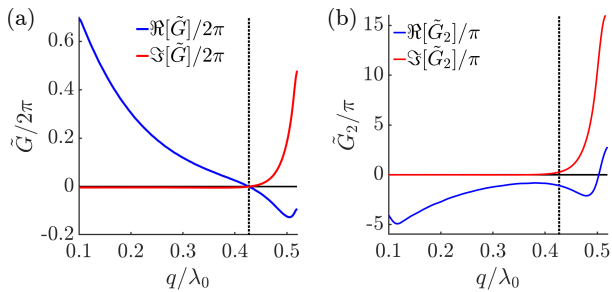}
\caption{\label{fig:fig_4} (a) Dispersive and dissipative parts of $\tilde{G}/2\pi$ for a DPhoC with a lattice period $0.5575\lambda_0$ and air-hole radius of $0.092\lambda_0$; this corresponds to the red square in Fig.~\ref{fig:fig1}(b). (b) Real and imaginary parts of $\tilde{G}_2/2\pi=\frac{1}{2}\frac{d^2 f_c/f_0}{dq^2/\lambda_0^2}$. The black dotted line gives the location of the BIC at $q=0.43\lambda_0$.}
\end{figure}

\emph{Estimated optomechanical coupling strengths}.--- Since a BIC has no radiative loss, its decay rate is given by some intrinsic loss. To gauge the achievable optomechanics performance of the DPhoC, we must estimate the internal loss channels of the PhC slab. In the following, we explore the DPhoC's performance for a set of realistic, albeit challenging, experimental parameters. In appendix \ref{appendix_6}, we also explore a more readily attainable parameter set. We consider intrinsic loss governed by material absorption and use experimental studies of GaAs microdisks~\cite{Michael2007} to obtain $\Im[n] = 4.4 \times 10^{-6}$, see appendix \ref{appendix_6} for details. This allows us to estimate the \emph{lower bound} on the achievable cavity decay rate, assuming that disorder-related loss and finite-size effects of both the beam and the sample can be ignored. We discuss these effects further in appendix \ref{appendix_6}, where it is shown that the DPhoC is surprisingly immune from finite-size effects. We also note that ultrashort Fano cavities have been shown to suffer less from finite-waist effects~\cite{Naesby2018}, illustrating a further advantage of working with compact cavities using BICs.\\
\indent To highlight the large single-photon cooperativity achievable with the DPhoC, we now change the lattice parameters to boost the dispersive linear coupling at the BIC location: a period of $0.7\lambda_0$, radius of $0.27\lambda_0$, and thickness of \unit{100}{\nano\meter}, indicated by the red circle in Fig.~\ref{fig:fig1}(b). This system is very practical, with a double-slab structure very close to these parameters already demonstrated~\cite{Manjeshwar2020}. The transmittance map, shown in Fig.~\ref{fig:fig3}(a), exhibits a BIC located in the intermediate zone at $q=0.44\lambda_0$ (pink dotted line), shown explicitly by a sharp peak in the Q-factor in Fig.~\ref{fig:fig3}(b). The Q-factor has a maximum around $Q=\Re[f_c]/(2 \kappa) = 6.8 \times 10^5$, which is limited by material absorption, and is similar in magnitude with the highest Q-factors for a BIC reported to date~\cite{Jin2019}. In Fig.~\ref{fig:fig3}(c) we plot $G$ and find $G/2\pi = \unit{-46}{\giga\hertz\per\nano\meter}$ and $\kappa/2\pi = \Im[f_c] = \unit{140} {\mega\hertz}$ at the BIC. Because our system is so compact, we can achieve coupling strengths of the order of tens-hundreds of GHz/nm. This is orders of magnitude larger than conventional out-of-plane systems~\cite{Thompson2008, Gartner2018} and comparable to values seen for in-plane geometries~\cite{Chan2011,Gavartin2011,Leijssen2017,guo2019feedback}. The DPhoC has the advantage that no outer cavity is necessary, as opposed to the MiM geometry \cite{Thompson2008} or multi-element optomechanics approach \cite{Xuereb2012}, considerably simplifying fabrication and operation.\\
\indent The DPhoC has the potential to access the regime of single-photon optomechanics \cite{Rabl2011,Nunnenkamp2011} by obtaining a large single-photon quantum cooperativity. Using realistic parameters of suspended PhC slabs \cite{Norte2016,Manjeshwar2020} with a mechanical frequency of $\Omega_m/2\pi=\unit{150}{\kilo\hertz}$ and associated effective mass $m_{\mathrm{eff}}=\unit{1}{\nano\gram}$ yields a single-photon optomechanical coupling strength of $g_0/2\pi=G\cdot x_0/2\pi \sim \unit{3.4 \times 10^5}{\hertz}$ and a considerable $g_0/\kappa$ ratio of $\sim 0.0025$ ($x_{0}=\sqrt{\hbar/2m_{\mathrm{eff}}\Omega_m}$). These estimated values place our system in the non-resolved sideband regime and firmly in the ultrastrong single-photon coupling regime with $g_0/\Omega_m\sim 2.3$, complementing previous works~\cite{Leijssen2017,fogliano2019cavity}.  Further, assuming a realistically achievable mechanical Q-factor of $Q_m \sim 10^8$~\cite{Norte2016,Tsaturyan2017} yields a single-photon cooperativity \cite{Aspelmeyer2014,borkje2020critical} of $\mathcal{C} = 4g_0^2 Q_m/(\kappa \Omega_m) \sim 2.2 \times 10^6$, which is similar to Ref.~\cite{fogliano2019cavity} and three orders of magnitude larger than in Refs.~\cite{Leijssen2017,guo2019feedback}. When operating the device at moderate cryogenic temperatures ($T=4K$), we predict a remarkable \textit{single-photon} quantum cooperativity of $\mathcal{C}_q=\mathcal{C}/n_\mathrm{bath} \sim 4.0$ ($n_\mathrm{bath}=k_BT/\hbar\Omega_m$). A value exceeding unity has not been achieved in any cavity optomechanics system before. Thus, the DPhoC offers a promising alternative to proposals in the microwave domain~\cite{Via2015,romero2018quantum} or to cavity optomechanics with atoms~\cite{Murch2008,Brennecke2008}. \\
\indent At the BIC separation, the optomechanical coupling is purely dispersive. Isolating purely \emph{dissipative} coupling is also interesting for certain quantum protocols~\cite{Elste2009,Xuereb2011,Wu2014}. To this end, we look at the region around $q=0.29\lambda_0$ where $\Re[G] \sim 0$, indicated by the green line in Fig.~\ref{fig:fig3}(a). Here, the DPhoC exhibits a large dissipative coupling: $\Im[G] = \unit{12}{\giga\hertz\per\nano\meter}$. As we are far from the BIC condition, a large decay rate is found. However, dissipative coupling can be utilized for optomechanical cooling without the need for the "good cavity" limit~\cite{Elste2009}. \\
\indent Through tuning of the lattice parameters, it is also possible to place the lowest-order BIC at a point of pure quadratic coupling by shifting it to where $\Re[G]$ vanishes, see Fig.~\ref{fig:fig_4}(a). The second derivative, $G_2$, is shown in Fig.~\ref{fig:fig_4}(b), illustrating that the quadratic coupling is finite where the linear coupling vanishes. We find a coupling of $\Re[G_2]/\pi = \unit{87}{\mega\hertz\per\nano\meter\squared}$ (and a dissipative coupling of $\Im[G_2]/\pi = \unit{6.7}{\mega\hertz\per\nano\meter\squared}$) for a $\kappa/2\pi=\unit{210}{\mega\hertz}$. This compares well with the values of $\unit{4.5 \rightarrow 30}{\mega\hertz\per\nano\meter\squared}$ reported by Sankey \emph{et al.}~\cite{Sankey2010}, but the DPhoC has the advantage of being many orders of magnitude more compact, and relies on a different mechanism of evanescent coupling rather than radiation pressure. It remains an open question whether the DPhoC system can be optimized to reach the values of $G_2 \sim \unit{1}{\tera\hertz\per\nano\meter\squared}$ reported for state-of-the-art planar PhC cavities~\cite{Paraiso2015}.\\
\indent \emph{Conclusions}.--- Combining light propagation in both free-space and guided-mode form, the DPhoC system merges the strengths offered by in-plane and out-of-plane optomechanical systems. We have estimated linear optomechanical coupling rates orders of magnitude larger than conventional end-mirror and MiM platforms, at moderately low optical decay rates, potentially leading to a single-photon quantum cooperativity exceeding unity. The DPhoC constitutes a versatile optomechanics platform able to access different regimes of optomechanical coupling that can be used to explore various quantum protocols in the non-resolved sideband regime~\cite{wilson2015measurement,Rossi2018,gut2019stationary,guo2019feedback,lau2020ground}, in particular in the framework of pulsed optomechanics~\cite{vanner2011pulsed,vanner2011selective,clarke2020generating} or frequency-dependent mirrors \cite{Naesby2018,Vcernotik2019}. For instance, the strong frequency dependence of the DPhoC's mirrors can be exploited in optomechanical cooling, as recently suggested in Ref.~\cite{Vcernotik2019}. The geometries described here represent a proof of concept and we expect optimized structures to yield even better performance. We envision many potential pathways from this work, including: squeezing of the guided-mode resonance in space using a defect cavity on the PhC slabs to boost photon-phonon co-localization~\cite{notomi2006optomechanical}, or utilizing phononic BICs \cite{zhao2019mechanical} alongside their photonic counterparts.

\begin{acknowledgments}
This work was partially supported by Chalmers’ Excellence Initiative Nano, the Knut and Alice Wallenberg Foundation through the Wallenberg Center for Quantum Technology (WACQT), the Swedish Research Council, and the QuantERA project C'MON-QSENS! Some of the numerical calculations were performed on resources provided by the Swedish National Infrastructure for Computing at C3SE.
\end{acknowledgments}

\appendix

\section{Numerical Methods}\label{appendix_1}
The numerical calculations are a combination of simulations based on the finite-element frequency domain method (using COMSOL Multiphysics) and rigorous coupled-wave analysis (RCWA) (using the $S^4$ code \cite{Liu20122233}). Where possible, results were obtained using both methods and excellent agreement to within a few percent was found.

\section{Fano Resonance}\label{appendix_2}
The interference between the direct transmission of light and the guided mode of a structured slab leads to unity reflection near the guided-mode resonance, $\omega_F$, with a width $\kappa_e$. Due to the large Q-factor of the underlying guided modes, Fano resonances are well described by CMT applied to a single resonator with two ports \cite{Haus1984,Fan2003}. Both $\omega_F$ and $\kappa_e$ can be found by calculating the reflection or transmission spectrum using numerical techniques to solve Maxwell's equations and fitting the following expressions:
\begin{eqnarray}
r(\omega) =  \frac{r_d(\omega-\omega_F) + t_d \kappa_e}{(\omega-\omega_F) + i\kappa_e} \\
it(\omega) =  \frac{-ir_d \kappa_e +  i t_d (\omega-\omega_F)}{(\omega-\omega_F) + i\kappa_e}, \label{eq_fano_fit}
\end{eqnarray}
which are derived under the assumption that the system possesses time-reversal symmetry, conservation of energy, and even symmetry with respect to the mirror plane. $r_d$ and $t_d$ are given by the reflectivity and transmission of a homogeneous slab with an effective refractive index \cite{Fan2003}. For a structured slab with air holes of radius $a$ and period $\Lambda$, the effective index is given by $n_{\mathrm{eff}} = (1-\eta)n + \eta$ where $\eta = \pi a^2/\Lambda^2$. An example of the fit is shown by the red-dashed line in Fig.~1(c). The radiative decay is quantified by $\kappa_e$ and linked to the width of the Fano lineshape (given by the shaded blue region in the plot). It describes the in- and out-coupling of the guided mode to external radiative channels. The inverse of $\kappa_e$ gives the typical travel time of a photon within the slab. Smaller air holes lead to a reduced $\kappa_e$ due to decreased scattering of the in-plane light, but this comes at the price of a larger impact from internal loss \cite{Suh2003}, measured by $\kappa_i$, which includes the impacts of various loss channels such as material loss and lattice imperfections.

\section{Bound state in the continuum theory}\label{appendix_3}
\begin{figure}[t!]
\includegraphics[width=8cm]{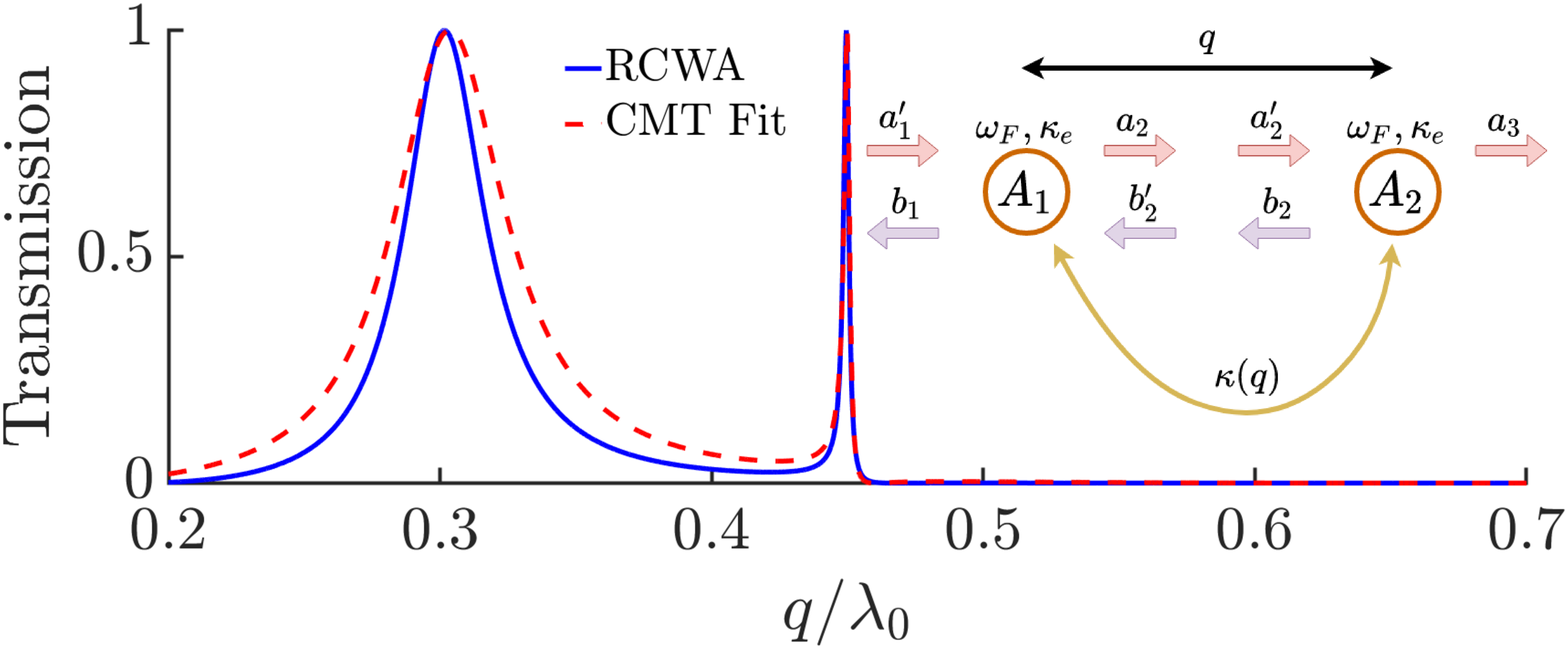}
\caption{\label{fig:fig_supp1}  Slice of the transmittance for $f=0.99 f_0$ calculated with RCWA (blue line) for a DPhoC with $100$-nm-thick slabs, $\Lambda = 0.7 \lambda_0$ and $a=0.27 \lambda_0$. A fit of the CMT (red-dashed line) is shown, along with an inset illustrating the CMT.} 
\end{figure}

The DPhoC is modelled as two resonators within CMT, $A_1$ and $A_2$, which obey the following two coupled first-order differential equations~\cite{Suh2005}
\begin{widetext}
\begin{eqnarray}
\partial_t A_1(t) = \left(-i\omega_F -  \kappa_e \right)A_1(t) + \sqrt{-\kappa_{e} (r_d+it_d)} (a_1 + b_2 e^{ikq}) + i\zeta A_2(t) \nonumber \\
\partial_t A_2(t) = \left(-i\omega_F -  \kappa_e \right)A_2(t) + \sqrt{-\kappa_{e} (r_d+it_d)} a_2e^{ikq} + i\zeta A_1(t), \label{eq_fano_double_slab}
\end{eqnarray}
\end{widetext}
where evanescent coupling is described by the real parameter $\zeta = C e^{-q/\delta}$ and coupling via photon propagation is described by the complex term $e^{ikq}$. $a_1, b_1,a_2, b_2$ and $a_3$ are the incoming and outgoing field amplitudes on either side of the slabs, and are defined in the inset of Fig.~\ref{fig:fig_supp1}. The transmission, $it=a_3/a_1$, can be found by Fourier transforming, and the remaining parameter, $\zeta$, is found by fitting the spectrum for fixed frequency and variable $q$. An example of this fitting procedure is shown in Fig.~\ref{fig:fig_supp1} by the red-dashed line. The fit is excellent, showing that this simple model captures both near- and far-field coupling between the slabs.

To illustrate why BICs occur for the DPhoC, we will make some drastic simplifications to the CMT that reveal the essential mechanisms more clearly. The physics we are interested in depends on the interaction of the resonances in each slab and not the direct process which is controlled by $r_d$ and $t_d$, therefore we set them to zero. This ``flat-background'' approximation is most valid for PhC slabs with large air holes and, hence, a lower effective refractive index. The coupling of the two resonator modes leads to hybridization into even and odd ``super-modes''~\cite{Song2009}:
\begin{equation}
A_{\mathrm{even/odd}}(t) = \frac{A_1(t) \pm A_2(t)}{\sqrt{2}}, 
\end{equation}
which have the following energies and decay rates
\begin{eqnarray}
\omega_{\mathrm{even}/\mathrm{odd}} = \omega_F \mp \left( \zeta(q) - \kappa_{e} \sin(k_0q) \right),\\
 \gamma_{\mathrm{even}/\mathrm{odd}} = \kappa_e + \kappa_i \pm \kappa_e \cos(k_0q),
\end{eqnarray}
where we have made the approximation that the right-hand side can be evaluated at the Fano energy, $\omega_F = c k_0$. The mode frequency shows a splitting between the even (which is at a lower energy) and the odd mode about the Fano energy, with contributions from both the near-field and far-field coupling, just as we observed in Fig.~2(a). These equations also reveal the presence of BICs: for no internal loss, coupling to output channels vanishes for $ \cos(k_0q) = \mp 1 $, which is just the usual FP resonance condition and reveals an infinite number of such BICs. Interestingly, to this approximation, the near-field coupling does not affect the BIC location; the BICs we observe are predominately a far-field phenomena. 

\begin{figure}[t!]
\includegraphics[width=8cm]{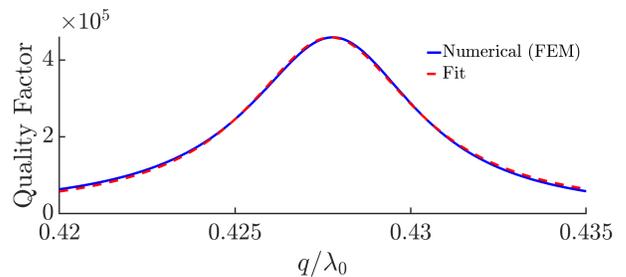}
\caption{\label{fig:fig_supp3} Quality factor of the cavity modes around a BIC for two PhC slabs of $100$ nm thickness, period of $0.5575\lambda_0$ and $a=0.092\lambda_0$. Shown also with a Lorentzian fit is $Q_{\mathrm{fit}}(q) = C^2/((q - q_0)^2)$, where $C = 1.9809  \lambda_0^2 $ and $q_0 = 0.43 \lambda_0$.} 
\end{figure}

Further evidence that we are indeed observing BICs comes from the quadratic dependence of $1/Q$ on $q-q_0$, where $q_0$ is the slab separation corresponding to a BIC~\cite{Blanchard2016}. This is confirmed by fitting $Q(q-q_0)$ and is shown in Fig.~\ref{fig:fig_supp3}.

\section{The near-field region}\label{appendix_4}
\begin{figure}[b!]
\includegraphics[width=8cm]{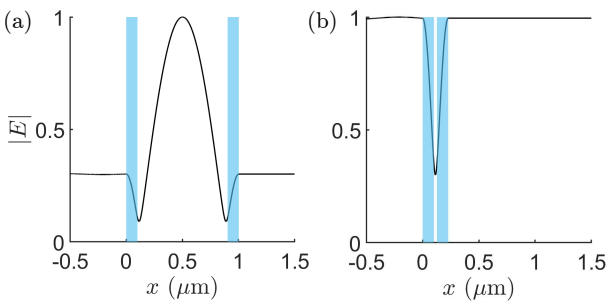}
\caption{\label{fig:fig_supp_near_field} (a) The lowest-order Fabry-P\'erot electric field profile and (b) the high-transmittance near-field mode. Calculated using the transfer-matrix method for incident light from the left-hand side at a wavelength of \unit{1550}{\nano\meter}.} 
\end{figure}

Here we discuss the near-field region shown in Figs.~2(a) and 2(b), where photon propagation between the slabs is negligible and evanescent coupling dominates. For the structured slabs, this region is indicated by the eigenmodes deviating from the bands of high transmittance. The eigenmodes become very lossy ($Q \sim 10$) and so are not shown in Fig.~2(a) for clarity. These modes could be useful for cavity optomechanics if we borrow the MiM philosophy and the DPhoC was placed within a larger cavity to recycle the leaked light.\\
\indent We also observe an interesting high-transmittance branch for the homogeneous slabs in Fig.~2(b). It derives from a family of leaky modes which do not correspond to FP modes. This is illustrated nicely in Fig.~\ref{fig:fig_supp_near_field}, where the electric field profile of the lowest-order FP mode and the near-field-zone mode are compared; the field of the former is concentrated within the cavity between the slabs, and the field of the latter is concentrated much more within the slabs. As the slabs are not structured and the incoming light is normally incident, it cannot be a consequence of near-field coupling and instead we speculate that it is similar in nature to zero-frequency modes seen for single slabs \cite{Llorens2014,Ismail2016}. Note that we do not include Casimir forces, which are derived from vacuum quantum fluctuations and are not present in our classical calculations.

\begin{table*}\begin{ruledtabular}\begin{tabular}{lllllllll}
$L$ & $\kappa/2\pi$ (MHz) & $G/2\pi$ (GHz/nm) & $g_0/2\pi$ (kHz) & $g_0/\kappa$ & $g_0/\Omega_m$ & $\kappa/\Omega_m$ & $\mathcal{C}$ & $\mathcal{C}_q$\\
17\,$\mu$m & 8.8 & 11.4 & 7.3 & $8.3 \times 10^{-4}$ & $1.5 \times 10^{-2}$ & 18 & 49 & $2.9\times 10^{-4}$\\
775\,nm & 193 & 250 & 161 & $8.3 \times 10^{-4}$ & $0.32$ & $390$ & 1100 & $6.4 \times 10^{-3}$\\
\end{tabular}\end{ruledtabular} \caption{\label{tab:table1} Optomechanical parameters for optical microcavities of length \unit{17}{\micro\meter} and $\lambda_0/2$. Common parameters between both sets are a finesse of $500000$, $m_{\mathrm{eff}}= \unit{40}{\nano\gram}$, $\Omega_m/2\pi = \unit{500}{\kilo\hertz}$, and $Q_m = 10^6$ at $T = \unit{4}{\kelvin}$.}\end{table*}

\section{Comparison of the DPhoC to a Fabry-P\'erot-type optomechanical microcavity}\label{appendix_5}

For the end-mirror configuration of length $q$, the linear dispersive coupling rate is given by $\Re[g_0] = \omega_c x_{0}/q$, where $x_0=\sqrt{\hbar/2m_{\mathrm{eff}}\Omega_m}$ is the zero-point motion and $\omega_c$ is the cavity frequency. For the MiM geometry, the maximum linear coupling rate is $2|r|$ times larger than the corresponding end-mirror geometry of the same total cavity length, where $|r|$ is the reflectivity of the inner membrane. In principle, $g_0$ can be increased as we decrease the length down to $q = \lambda/2$ (below which no FP resonance is supported). The decay rate of an end-mirror cavity is given by $\kappa_c = \frac{\pi c}{2q \mathcal{F}}$, where $\mathcal{F}$ is the cavity finesse. This means that $\Re[g_0]/\kappa_c$ is independent of length.\\
\indent Let us estimate the optomechanical parameter regime achievable with a FP-type optomechanical microcavity, which in turn allows us to compare to the performance of the DPhoC. To this end, we combine parameters from independent realizations of state-of-the-art optical microcavities~\cite{wachter2019silicon} and distributed Bragg reflector (DBR)-based high-reflectivity mechanical resonators~\cite{groblacher2009demonstration,cole2011phonon,weaver2016nested} in order to obtain an estimate on the potential of a FP-based optomechanical microcavity. We consider an optical microcavity of length $q=\unit{17}{\micro\meter}$ with a finesse of $\mathcal{F}=5 \times 10^5$ at telecom wavelengths, which has been recently realized in chip-based silicon microcavity arrays~\cite{wachter2019silicon}. Note that a slightly smaller finesse of $1.8 \times 10^5$ has been achieved in a 5-cm-long FP-based optomechanical system~\cite{weaver2016nested}. Both of these cavities employed multilayer coatings, i.e., DBRs,  to achieve such an exceptionally large finesse. Hence, the mechanical resonator has to be realized via a suspended DBR~\cite{cole2011phonon} or a DBR on a mechanical resonator~\cite{groblacher2009demonstration,weaver2016nested} to obtain such high finesse values. These systems have typical mechanical parameters of $\Omega_m/2\pi\sim \unit{500}{\kilo\hertz}$, $m_{\mathrm{eff}}\sim \unit{40}{\nano\gram}$ and a mechanical quality factor $Q_m\sim 10^6$ at $T\sim \unit{4}{\kelvin}$ \cite{groblacher2009demonstration,cole2011phonon,weaver2016nested}. Note that the DBR limits the performance of the mechanical resonator, in particular, resulting in a lower mechanical quality factor and larger effective mass compared to state-of-the-art DBR-free mechanical resonators, which routinely achieve values of $m_{\mathrm{eff}}\sim \unit{1}{\nano\gram}$, $Q_m>10^8$. All together, this leads to the set of parameters displayed in Table~\ref{tab:table1} and, hence, to much less advantageous optomechanical values than the DPhoC we propose in this work, with the exception of a slightly improved $\kappa/\Omega_m$ ratio.\\
\indent Also shown in Table~\ref{tab:table1} are the parameters for the minimal cavity length of $q=\lambda_0/2$ of such a hypothetical FP cavity. Despite this microcavity having a larger $G$ than the DPhoC we consider, such a system suffers from the ratio $g_0/\kappa$ being independent of cavity length, and the worse performance of the mechanical resonator compared to PhC-based mechanical resonators. Hence, we conclude that a FP microcavity will lead to a worse performance than a DPhoC system.

\section{Estimation of loss channels}\label{appendix_6}

\subsection{Material absorption loss}
To estimate the ultimate upper bounds on the BIC's Q-factor, we need an estimate of the intrinsic material loss. We extracted material-based absorption for GaAs using Ref.~\cite{Michael2007}, where a loss rate of $\frac{\kappa_i}{2\pi} \sim \unit{0.5}{\giga\hertz}$ was measured for GaAs microdisks at \unit{1600}{\nano\meter}. This yields an absorption coefficient $\alpha=\kappa_i/v_g \sim \unit{0.3}{\centi\reciprocal\meter}$ with the group velocity, $v_g$, estimated as $\sim10^8$ m/s. For the 100-nm-thick membranes we consider in this work, we get a material absorption of about $3\,$ppm; this is an overestimation of the loss as some of the electric field of the mode will be concentrated in the air-holes rather than the GaAs. The imaginary component of the refractive index can then be found from: $\Im[n] = \frac{\alpha \lambda_0}{4 \pi}$~\cite{Xu2009}. For our operation wavelength of $\lambda_0=1550$ nm, this gives $\Im[n] \sim 4.4 \times 10^{-6}$. This value, along with a mechanical quality factor of $Q_m = 10^8$, will be denoted as parameter set I and displayed in Table~\ref{tab:table2}. This set was used earlier and represents challenging, but achievable, parameters that are state-of-the-art in both mechanics and photonics. Using this value of $\Im[n]$ for a single PhC slab gives a max reflectance of $R = 0.99998$ near the Fano resonance.

\begin{table*}\begin{ruledtabular}\begin{tabular}{llllllll}
Set &  $Q_m$ & $\Im[n]$ & $R$ & $\kappa/2\pi \ ({\mega\hertz})$ & $g_0/\kappa$ & $\mathcal{C}$ & $\mathcal{C}_q$\\
(I) &  $10^8$ & $4.4 \times 10^{-6}$ & $0.99998$ & $140$ & $0.0025$ & $2.2\times 10^6$ & $4.0$\\
(II) &  $10^7$ & $2 \times 10^{-4}$ & $0.999$ & $6200 $ & $5.5\times 10^{-5}$ & $5000$ & $0.009$\\
\end{tabular}\end{ruledtabular} \caption{\label{tab:table2} Optomechanical parameters for linear dispersive coupling for the DPhoC. For both parameter sets $ \Omega_m/2\pi=\unit{150}{\kilo\hertz}$, $G/2\pi = \unit{-46}{\giga\hertz\per\nano\meter}$ and $m = \unit{1}{\nano\gram}$.}\end{table*}

For set II, we estimate the corresponding effective $\Im[n]$ for a max reflectance of $0.999$, which was achieved in a single \ch{Si_3N_4} PhC slab in Ref.~\cite{Chen2017}. This yields $\Im[n]=2 \times 10^{-4}$. This is not entirely appropriate as the devices in the aforementioned reference were limited by scattering rather than material absorption, but it gives an indication of the effects of non-unity reflectance and the resulting optical Q-factor of $\sim 10^4$ is in line with values found for typical BIC systems~\cite{lee2014fabricating}. The parameters for set II are also shown in Table~\ref{tab:table2}.

\subsection{Transverse effects}
In our estimation of $g_0$, we have ignored loss from transverse effects such as wavefront curvature, non-perfectly-parallel mirrors, and finite-area structures. These unavoidable limitations are a consequence of incident light coupling into modes located over a finite region of k-space, leading to additional loss channels. Relevant to our discussions is that ultrashort cavities built from Fano mirrors have been shown to suffer less from finite-waist effects \cite{Naesby2018}. There is also the possibility of designing Fano mirrors with focusing abilities \cite{Fattal2010,Guo2017}. Furthermore, resonance-trapped BICs have been shown to display a large Q-factor over a wide range in k-space \cite{Kodigala2017}, and recently the merging of multiple BICs has been used to suppress out-of-plane scattering losses~\cite{Jin2019}.

\begin{figure}[b!]
\includegraphics[width=8cm]{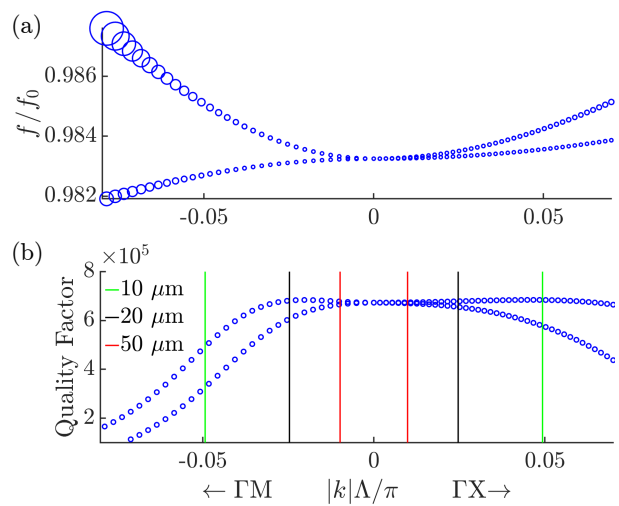}
\caption{\label{fig:fig_supp_bandstructure} (a)~Energy splitting and (b)~quality factor of the DPhoC eigenmodes around the $\Gamma$ point. DPhoC parameters are: lattice period of $0.7\lambda_0$, air-hole radius of $0.27\lambda_0$, and slab separation of $0.439 \lambda_0$.} 
\end{figure}

To demonstrate that the DPhoC is surprisingly immune from finite-size effects, we have calculated the Q-factor for wavevectors away from the high-symmetry $\Gamma$ point of the first Brillouin zone for a square lattice (see Fig.~\ref{fig:fig_supp_bandstructure}). To save simulation time, we explore slices in k-space in the direction from $\Gamma$ to the other high-symmetry points $X$ and $M$. A detailed calculation would integrate over a specified area of k-space---a thorough discussion of including finite-beam-waist size effects in reflection and transmission spectra can be found in the supplementary material of Ref.~\cite{Manjeshwar2020}. We observe that the mode at $\lambda_0$ is doubly degenerate at the $\Gamma$ point and splits in energy, which can be seen in Fig.~\ref{fig:fig_supp_bandstructure}(a). Importantly, the Q-factor remains well above $10^5$ in a large region of k-space; this is shown in Fig.~\ref{fig:fig_supp_bandstructure}(b). In realistic devices, there is a compromise between the lateral size of the device, which will affect mechanical properties, and the beam-waist size to achieve maximum slab reflectivities. A Gaussian beam can be represented as a sum of angled plane-wave components weighted by a Gaussian distribution with a standard deviation given by the beam divergence, $\theta_0=\lambda/(\pi w_0)$, where $w_0$ is the beam-waist size. In Fig.~\ref{fig:fig_supp_bandstructure}(b), we represent the beam divergence for beam-waists of 10, 20 and \unit{50}{\micro\meter}, which are typical values used in experiments~\cite{Norte2016, Moura2018, Gartner2018}.

\subsection{Estimate for SiN-based system}
The physics discussed throughout this work is not material dependent and can be expected to be applicable for SiN-based systems~\cite{Norte2016,Bernard2016,Chen2017,Nair2017,Piergentili2018,Gartner2018}, which are more commonly used in optomechanics. Here, we provide an estimate of the optomechanical parameters obtainable with a such systems. The optical Q-factor of the BIC can be approximated using~\cite{Xu2009}, $\frac{1}{Q_{\mathrm{BIC}}}  = \frac{1}{Q_{\mathrm{abs}}}  \sim \frac{2 \Im[n]}{ \Re[n]}$, which we have confirmed to be an accurate estimate for GaAs PhC slabs. $\Im[n]$ as small as $2 \times 10^{-6}$ has been measured for membranes of close to \unit{100}{\nano\meter} thickness at a wavelength of \unit{1064}{\nano\meter}~\cite{serra2016microfabrication}. As $\Re[n] = 2.021$, this gives $Q_{\mathrm{BIC}} \sim 5 \times 10^5$. Assuming similar optomechanical coupling rates and mechanical properties, then the ultimate achievable optomechanical parameters of a SiN-based DPhoC should be similar to set I in Table~\ref{tab:table2}.

\bibliographystyle{apsrev4-2}
\bibliography{bib}

\end{document}